\documentclass[aip,pop,preprint]{revtex4-1}
\usepackage{graphicx}%
\usepackage{dcolumn}%
\usepackage{bm}%
\usepackage{amsmath}%
\begin{document}
\title{Instabilities and Turbulence in Low-$\beta$ Guide Field Reconnection Exhausts with Kinetic Riemann Simulations}
\date{\today}
\author{Qile Zhang}
\author{J. F. Drake}
\author{M. Swisdak}
\affiliation{University of Maryland, College Park, Maryland, 20742, USA}
\begin{abstract}
The role of turbulence in low-$\beta$, guide-field reconnection
exhausts is explored in 2D reconnection and 2D and 3D Riemann
simulations. The structure of the exhaust and associated turbulence is
controlled by a pair of rotational discontinuities (RDs) at the
exhaust boundary and a pair of slow shocks (SSs) that are generated by
counterstreaming ions beams. In 2D the exhaust develops
large-amplitude striations at the ion Larmor radius scale that
are produced by electron-beam-driven ion cyclotron waves. The electron
beams driving the instability are injected into the exhaust from one
of the RDs. However, in 3D Riemann simulations, the additional
dimension (in the out-of-plane direction) results in strong Buneman
and electron-electron streaming instabilities at the RD which suppress
electron beam formation and therefore the striations in the
exhaust. The strength of the streaming instabilities at the RD are
controlled by the ratio of the electron thermal speed to Alfv\'en
speed, lower thermal speed being more unstable. In the 3D simulations
an ion-ion streaming instability acts to partially thermalize the counterstreaming
ion beams at the SSs. This instability is controlled by the ratio of
the sound speed to Alfv\'en speed and is expected to be stable in the
low $\beta$ solar corona. The results suggest that in a guide field
reconnection exhaust with $1 \gg \beta > m_e/m_i$, the kinetic-scale
turbulence that develops will be too weak to play a significant role
in energy conversion and particle acceleration. Therefore, the energy
conversion will be mostly controlled by laminar physics
or multi-x-line reconnection.

\end{abstract}

\maketitle

%\onecolumngrid
\section{Introduction}
Magnetic reconnection and turbulence are
fundamental processes in plasma systems. Reconnection
converts magnetic energy to plasma high speed flows, heating and
energetic particles through a change of magnetic
topology. Turbulence contributes to particle scattering, transport,
acceleration, energy dissipation and so on. These two phenomena could
intertwine, so it is of fundamental importance to understand the role of turbulence
in reconnection, especially in the process of energy conversion.

%Reconnection conversts magnetic energy into plasma energy and cause explosive phenomina like flares CME and geostorms, causing significant threat to human civilization on earth and in space. Therefore it is very motivated to understand the energy conversion mechanism. However, the existence of turbulence could modify the process of particle heating and acceleration and lead to different dominating mechanism. Therefore the role of turbulence in reconnection is essential. 

Here we focus our attention on turbulence in single x-line
reconnection rather than multi x-line reconnection. Turbulence is
often driven by instabilities. Previous observational and numerical
studies have investigated instabilities and turbulence in reconnection
near the diffusion region and along the magnetic separatrices that
emanate from the magnetic x-line
\cite{Drake2003,Cattell2005,Che2011,Graham2017,Price2016,Le2017,Price2017}
as well as in the exhaust downstream of the
x-line\cite{Liu2012,Munoz2016,Pucci2017,Eastwood2018,Jiansen2018}. They
could contribute to anomalous resistivity and viscosity, cause
dissipation and scattering and so on. The major region of energy
conversion is the reconnection exhaust downstream of the x-line.
%Recently He et al.\cite{Jiansen2018} and Pucci et al.\cite{Pucci2017} investigate turbulence in antiparallel reconnection exhausts. 
Recently, Eastwood et al.\cite{Eastwood2018} observed a guide field
reconnection exhaust $\sim$100 $d_i$ (with $d_i$ the ion inertial length) downstream of the x-line with a
$\beta$ of order unity and identified electron holes. Munoz et
al.\cite{Munoz2016} used low-$\beta$ ($<$0.1) 2D particle-in-cell
(PIC) simulations to explore the turbulence present in guide field
reconnection exhausts. However, the simulation domains only extended
about 6 $d_i$ downstream of the x-line so the turbulence further
downstream remains to be explored. In this paper, we will also focus
on the guide field and low-$\beta$ regime, which is relevant to the
solar corona and the inner heliosphere, where reconnection can drive
powerful releases of magnetic energy in solar flares and coronal mass
ejections (CMEs).
%recently Eastwood et al look at a guide field reconnection exhaust using MMS and observe turbulence and Near the diffusion region and along the separatrices, the turbulence could play a role in creating anomalous resistivity viscosity, dissipation, mixing particles. Yet most of the energy conversion happens in the reconnection exhaust where the field lines contract.  So it essential to know the role of turbulence in the exhausts, especially in the low$\beta$ environment applicable to the corona and inner heliosphere, where hot and energetic particle are constantly created by reconnection.

To study the exhaust, 2D PIC reconnection simulations are usually used
to capture kinetic effects. However, simulations of large systems at
low $\beta$ with sufficiently high ion-to-electron mass ratio are
computationally expensive so we employ the PIC Riemann simulation
model. The magnetic geometry of a Riemann simulation resembles a single
reconnection outflow downstream of the x-line without capturing the reconnection process near the x-line. A Riemann simulation reduces the dimension of an 2D outflow by neglecting the slow variation in the outflow
direction (x) so it becomes one dimensional (along y) in
its most basic formulation \cite{Lin1993}. The computational cost to carry out PIC Riemann simulations is therefore greatly reduced compared with PIC 2D reconnection simulations and we can consequently explore systems with a much larger spatial domain in y and z. However, to capture
kinetic-scale instabilities and turbulence, a small length in
the outflow (x) or guide field (z) direction can be
kept, resulting in 2D or 3D Riemann simulations. Previously 2D (x-y)
Riemann simulations have been used to model the instabilities and
turbulence of antiparallel reconnection exhausts using either
hybrid\cite{Scholer1998,Cremer1999,Cremer2000} or PIC
\cite{Liu2011(a),Liu2011(b)} models.

In an earlier paper we explored the structure of reconnection exhausts
in the low-$\beta$, strong-guide-field limit using PIC Riemann simulations\cite{Zhang2019}. We showed that the exhaust was bounded by RDs with parallel slow-shocks (SSs)
forming within the exhaust as expected from the magnetohydrodynamic
(MHD) model \cite{Lin1993}. This differs from the structure in the weak guide-field case, where the exhaust is bounded by Petschek-like switch-off slow shock/rotational discontinuity compound structures\cite{Liu2012,Innocenti2015,Innocenti2017}. The SSs in our simulations remained
laminar and were not effective in heating either electrons or ions
through the usual diffusive shock acceleration mechanism. In this
paper we explore in much greater depth the instabilities and
turbulence in low-$\beta$ guide field reconnection exhausts to reveal the parameter regimes in which the reconnection exhausts are laminar or turbulent.  We carry out a 2D reconnection simulation and 2D and 3D Riemann
simulations with various mass-ratios.

The guide field breaks the system's symmetry and leads to a density
cavity at the RD on one side of the exhaust (the side where the
electron velocity supporting the current across the RD points toward
the midplane of the exhaust) and a density bump at the other RD. In 2D
simulations the streaming electrons in the cavity of the RD drive the
Buneman instability at early time but later in time the dominant
instability moves to the core of the exhaust and leads to large
amplitude striations in the parallel current with
characteristic scales of the order of the ion Larmor radius $\rho_i$.  The
instability is an ion cyclotron wave driven by the strong electron
beam injected into the exhaust by the low density RD
\cite{Drummond1962}. However, in 3D simulations the Buneman instability as well as the electron-electron streaming instability
at the low density RD becomes much stronger (due to a non-zero $k_z$)
so that the electron beam injected into the exhaust becomes much
weaker and the development of the striations is suppressed. The result
points to the importance of full 3D simulations of reconnection to
understand the mechanisms for energy conversion. The strength of the streaming instabilities at the RD are
controlled by the ratio of the electron thermal speed to Alfv\'en
speed, with lower thermal speed being more unstable. The 3D simulations
also reveal the development of a weak ion-ion streaming instability
within the exhaust that did not appear in the 2D model. This instability partially thermalizes the counterstreaming
ion beams at the SSs. This weak instability is expected to be stabilized at lower beta where the ion relative drift (of order the Alfv\'en speed)  becomes larger than the sound speed. As a result, the counterstreaming ion beams that develop during low-$\beta$, guide field reconnection can propagate long distances without thermalizing. The conclusion,
therefore, is that in the regime $1 \gg \beta > m_e/m_i$ the instabilities and turbulence that develop are
expected to be weak in a realistic guide-field
reconnection exhaust. The dynamics of a single
x-line exhaust is therefore dominated by laminar processes in this regime\cite{Zhang2019}.

The organization of the paper is the following: in Sec.\ 2 we present
the results of simulations of 2D reconnection exhausts, including a
discussion of turbulence; in Sec.\ 3 we present the results of 2D
Riemann simulations and analyze the associated turbulence; in Sec.\ 4
we present the results of 3D Riemann simulations and show that there
are qualitative differences between 2D and 3D simulations and modest
versus large mass-ratio simulations; and in Sec.\ 5 we present the
conclusions and implications.

\section{Instabilities and turbulence in 2D reconnection exhausts
}
In this paper, we perform simulations using the particle-in-cell code {\tt p3d}\cite{Zeiler2002three‐dimensional}. The particles are advanced by the Newton-Lorentz equations of
motion and the fields by Maxwell's equations. We apply periodic boundary conditions to all boundaries, so we have two identical reconnection current sheets to achieve the periodic condition in y. However, we only focus on one current sheet in half of the domain. Here magnetic field strengths are normalized
to the initial asymptotic field in x direction $B_{\text{x,a}}$, densities to the initial asymptotic density $n_0$, lengths to the ion inertial length $d_i=c/\omega_{pi}$ based on $n_0$, times to the inverse ion cyclotron frequency $\Omega_{ci}^{-1}$, velocities to the Alfv\'en speed $C_{Ax}$ based on $B_{\text{x,a}}$ and $n_0$, and temperatures to $m_iC_{Ax}^2$. In this section, we show results from a 2D guide field reconnection simulation (Run 1) with a guide field equal to the reconnecting field ($B_{\text{z,a}}$=$B_{\text{x,a}}$), which is initialized by a force free configuration:
\begin{subequations}
\label{initallequations}
\begin{eqnarray}
B_x=B_{\text{x,a}}\ \tanh(y/w_0),\label{initequationa}
\\
 B_z=\sqrt{(B_{\text{x,a}}^2+B_{\text{z,a}}^2-B_x^2)},\label{initequationb}
 \\
 n=n_0,\label{initequationc}
 \end{eqnarray}
\end{subequations} 
Other parameters are shown in Table~\ref{table1}. This simulation has an initial total $\beta$ of 0.1. In this paper, the number of particles per grid cell (ppg) is 100 in 2D simulations and is 25 in 3D simulations. Besides, a $B_y$ perturbation highly localized in x is used to initiate a single x-line in the initial current sheet. The perturbation takes the form: $\delta B_y=\partial_x ((\sin(2\pi x/L_x))^{499}).$

\begin{figure*}
\includegraphics[width=\linewidth]{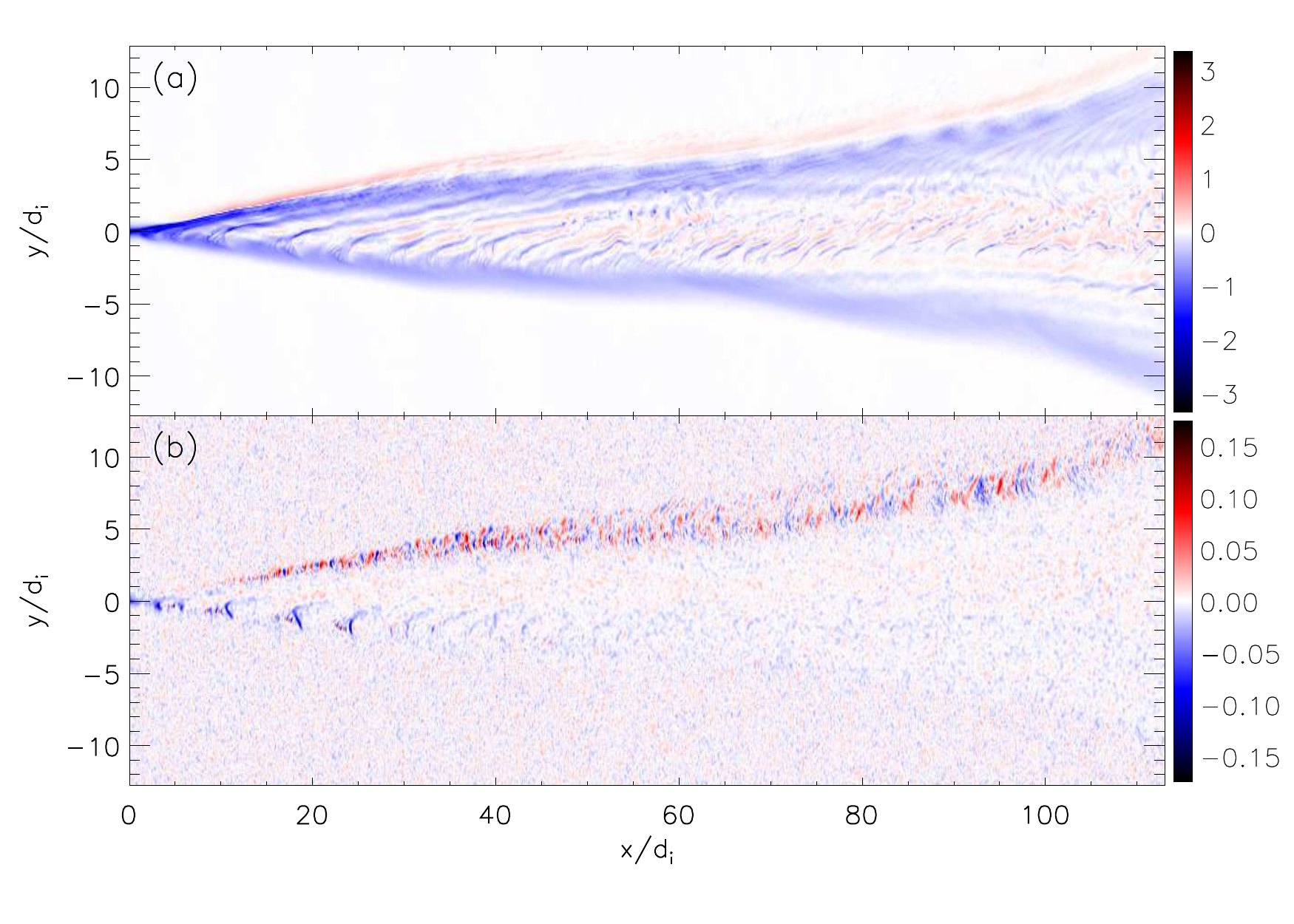}
\caption{\label{fig1} $J_\parallel$ (a) and $E_\parallel$ (b) of a reconnection exhaust in the 2D reconnection simulation Run 1}
\end{figure*}

This simulation creates a steady reconnection exhaust extending for
more than 100 $d_i$ along the outflow direction as shown in
Fig.~\ref{fig1}. This simulation was also analyzed in Zhang et
al.\cite{Zhang2019} with emphasis on the heating and the overall
structure. Here we focus on the instabilities and turbulence. In
Fig.~\ref{fig1}(a), we show the parallel current
$J_\parallel=\textbf{J}\cdot\textbf{B}/B$, which is mostly carried by
electrons. We see the two current sheets bounding the exhaust which
are the RDs\cite{Zhang2019}. Note that they are not the separatrices but are downstream of the separatrices since there is no change of magnetic topology across them. The core of the exhaust is highly
structured with oblique striations in the current. When these
striations first develop downstream of the x-line, the source is from
the low density RD. Further downstream, the core of the exhaust become
more turbulent and this turbulence appears to be disconnected from the
RD. In Fig.~\ref{fig1}(b), we show the parallel electric field
$E_\parallel=\textbf{E}\cdot\textbf{B}/B$, which indicates ongoing
instabilities at both RDs. Those at the bottom RD are strong close to
the x-line and fade away further downstream. The fluctuations
develop at short scale close to the x-line and then clump to larger
scales that link to the striations within the exhaust. The nature of
these instabilities and turbulence is dicussed in the next
section.

\section{Studies of instabilities and turbulence using 2D Riemann simulations
}
\subsection{Riemann simulations to explore reconnection exhausts}
To study the current striations and instabilities in the
reconnection exhaust in greater detail, we first simplify the
configuration by using Riemann simulations. As has been discussed previously
\cite{Lin1993,Scholer1998,Liu2012,Zhang2019}, Riemann simulations model the
reconnection exhaust by neglecting the dependence on the outflow
direction x. Here since we want to study the structures and
instabilities in the x-y plane as in the reconnection simulation, we keep
a long enough length in the x dimension to model the striations. Thus,
the 2D Riemann simulation is performed in a x-y domain. The parameters
(see Run 2 in Table~\ref{table1}) and initial profiles are almost the
same as Run 1 but there is a small initial magnetic field
$B_y=0.1B_{\text{x,a}}$ added so that the magnetic tension can drive
the outflow. In addition, the half width in y of the initial current
sheet $w_0$ is chosen to be close to the half width of the x-line
current sheet in Run 1 at the time of Fig.~\ref{fig1}. The Riemann
simulation result at time $t$ is a proxy of the reconnection exhaust
region at a distance $C_{Ax}*t$ downstream from the
x-line. Zhang et al.\cite{Zhang2019} demonstrated that Riemann
simulations capture the overall structure of the reconnection
exhaust. Here we show that they also capture the development of
turbulence and hence can be used as proxies for full simulations to
study that turbulence.

\begin{figure*}
\includegraphics[width=\linewidth]{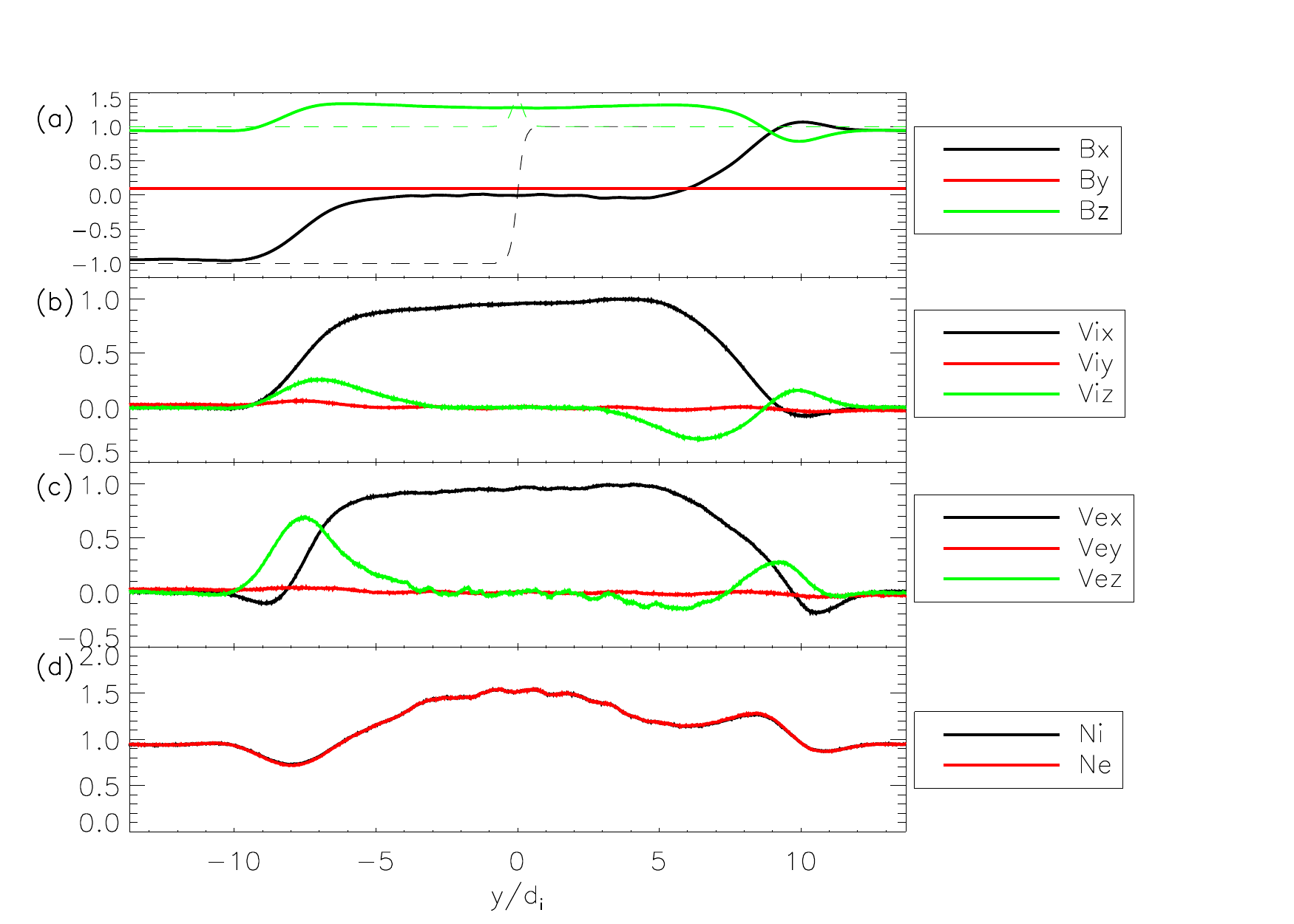}
\caption{\label{fig2} The overall 1D profiles (averaged over x) of the 2D Riemann simulation (Run 2). (a) the magnetic fields (with initial profiles in dashed lines of the same color), (b) ion velocities, (c) electron velocities and (d) ion and electron densities (which nearly overlap). }
\end{figure*}

We first show the overall 1D profiles (averaged over x) of this 2D
Riemann simulation in Fig.~\ref{fig2}. In panel (a), there are two
locations where the magnetic field rotates from the dominant x-z
direction to the z direction. These rotations form the RDs that bound
the exhaust. We also plot the initial magnetic field profiles in dashed lines for comparison. In panels (b) and (c), we present the three velocity
components of the electrons and ions.  The RDs drive the exhaust
velocity $V_x$ of both species. The RDs also drive out-of-plane plasma fluid flows
$V_{z}$ that are seen most easily in the ions. These flows are toward
the midplane of the exhaust and in the MHD model produce the SSs that
develop in guide field reconnection. The RDs are supported by a
current $J_z$ that produces the magnetic rotation. From Ampere's law
the direction of $J_z$ is the same (negative) at both RDs. The
electrons have a larger positive $V_z$ than the ions (to create a
negative $J_z$) so the electrons dominate the current at the RD.
Since $B_z$ and $B_y$ are both positive across the domain (panel
(a)), the electrons with positive $v_z$ at both RDs are flowing along
the field line towards the positive y direction. So the electrons that
carry the current at the left RD are at the same time accelerated into
the exhaust. The acceleration reduces the local density due to flux
continuity and creates a density cavity at the left RD (panel (d)). In
contrast, the electrons that carry the current at the right RD are
accelerated away from the exhaust, leading to a pileup of the
electrons and therefore a density increase. These density variations
at the RDs have no counterpart in MHD.

\begin{figure*}
\includegraphics[width=\linewidth]{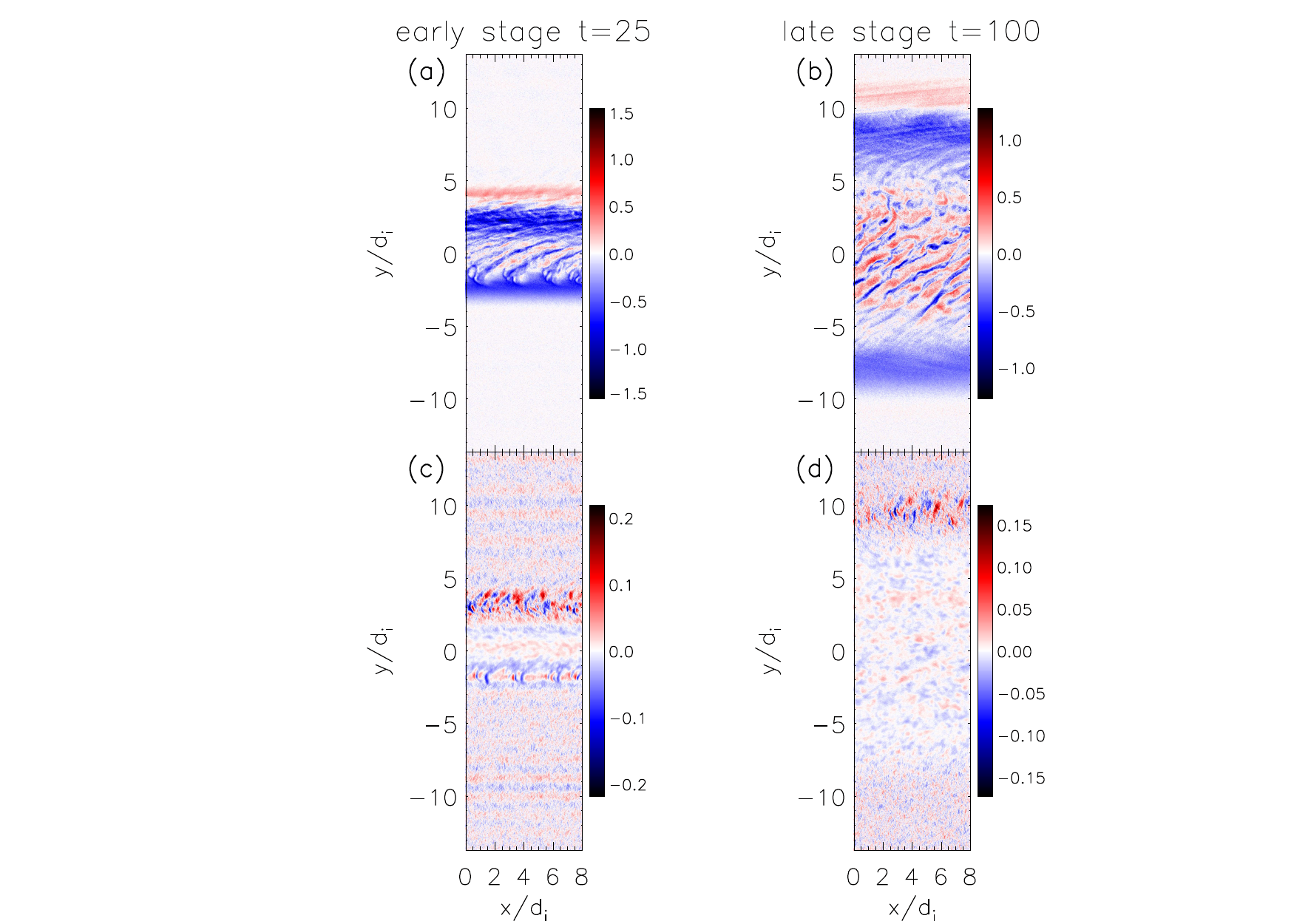}
\caption{\label{fig3} (a)(b) $J_\parallel$ and (c)(d) $E_\parallel$ at two times from the 2D Riemann simulation (Run 2).}
\end{figure*}

In Fig.~\ref{fig3}, we show the results from this 2D Riemann
simulation at both early and late time, as proxies of regions of
about 25 $d_i$ and 100 $d_i$ downstream of the x-line in the reconnection simulation in Fig.~\ref{fig1}. Although
$L_x=16$ in this simulation, we only show half the domain with
$x\in$[0,8] to facilitate the comparison to a 3D simulation of a
smaller domain shown later in Fig.~\ref{fig7}. In Fig.~\ref{fig3},
panels (a) and (b) show $J_\parallel$ and panels (c) and (d) show
$E_\parallel$. Comparing to Fig.~\ref{fig1}, we see that the Riemann
simulation captures all the essential features of the instabilities
and turbulence we discussed in the 2D reconnection simulation,
including the oblique current striations and the structuring of
$E_\parallel$. The amplitudes of these fluctuations are also
comparable to those in the reconnection simulation. This suggests that
the Riemann simulations can be used to explore the development of
turbulence in reconnection exhausts. Therefore, we will use
Riemann simulations to explore the nature of these instabilities
and the development of turbulence.

\subsection{Studies of instabilities and turbulence
} 
\begin{figure*}
\includegraphics[width=\linewidth]{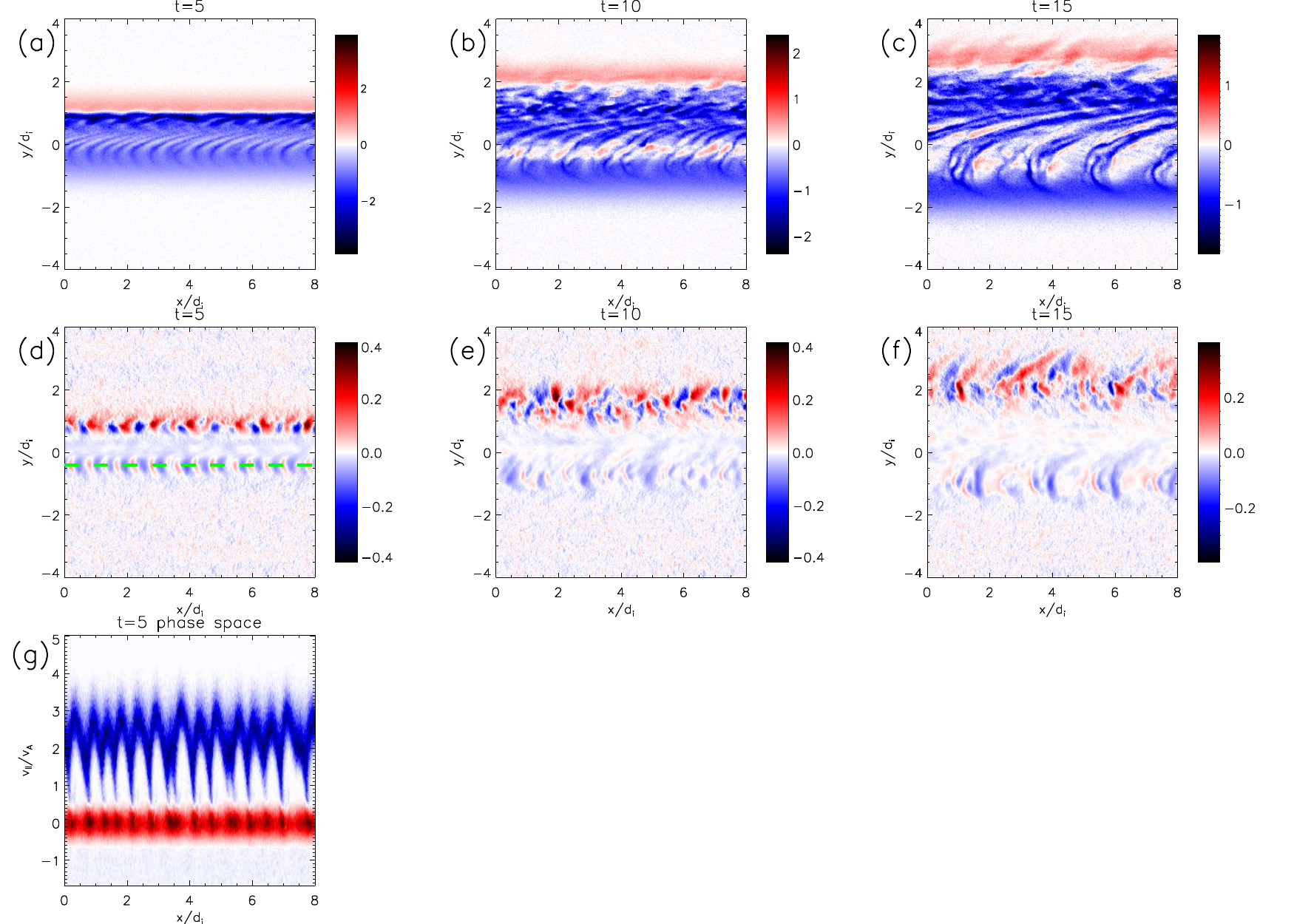}
\caption{\label{fig4} The early evolution from Run 2. In (a)-(c) $J_\parallel$  and in (d)-(f) $E_\parallel$. In (g) the phase space x-$V_\parallel$ which is taken along the green dotted line in (d). The ion population in red and the electron population in blue, with their maximum phase space densities normalized to be the same.}
\end{figure*}

In Fig.~\ref{fig4} we show the early evolution of the Riemann
simulation to understand the dynamics just downstream of the
x-line. Shown are the data from times $t=5$, $10$ and $15$. Panels
(a)-(c) show $J_\parallel$, and panels (d)-(f) show $E_\parallel$. In
panel (d) the instability in the bottom current sheet is still in, or
just past, its linear phase. Correspondingly, the current in panel (a)
is modified by the instability and the striations in the current begin
to extend from this current sheet into the exhaust. The reason why the
current sheet on this side affects the current structures in the
exhaust is that the current here is supported by electrons flowing
towards the midplane, as discussed in the last subsection. So the
structures of the current at this RD convect into the exhaust. Since
the exhaust outflow is in the positive x direction relative to the almost
stationary upstream plasma (Fig.~\ref{fig2}(c)), there is a velocity
shear that the electrons will experience when they flow from the
current sheet into the exhaust. That is why the striations tilt
towards the positive x direction. We also show the phase space
x-$V_\parallel$ in panel (g) which is taken along the green dotted
line in panel (d). The ion population is in red and the electron
population is in blue. Their maximum phase space densities are
normalized to be the same. Since both species are strongly modified by
the instabilty, we conclude that the instability is an oblique Buneman
instability. This instability can develop in this 2D system with a
wave vector $k_x$ because the current has an x component. The wave
vector in z can not develop in this 2D x-y simulation. As a cross
check, we verified that the dominant $k_x$ was consistent with the
prediction for the Buneman instability in Drake et
al.\cite{Drake2003},
\begin{equation}
    k=\delta \omega_{pe}/v_{de},
\end{equation}
where $\delta^2=(1+\sin^2\theta \omega_{pe}^2/\Omega_{ce}^2)^{-1}$.
By evaluating the local parameters: electron beam speed $v_{de}=2$,
wave vector direction relative to the field $\cos{\theta}=0.3$,
density $n=0.5$, magnetic field $B=1.4$, we calculate the wavelength
along x to be 0.6, which matches that found in the
simulation.

\begin{figure*}
\includegraphics[width=\linewidth]{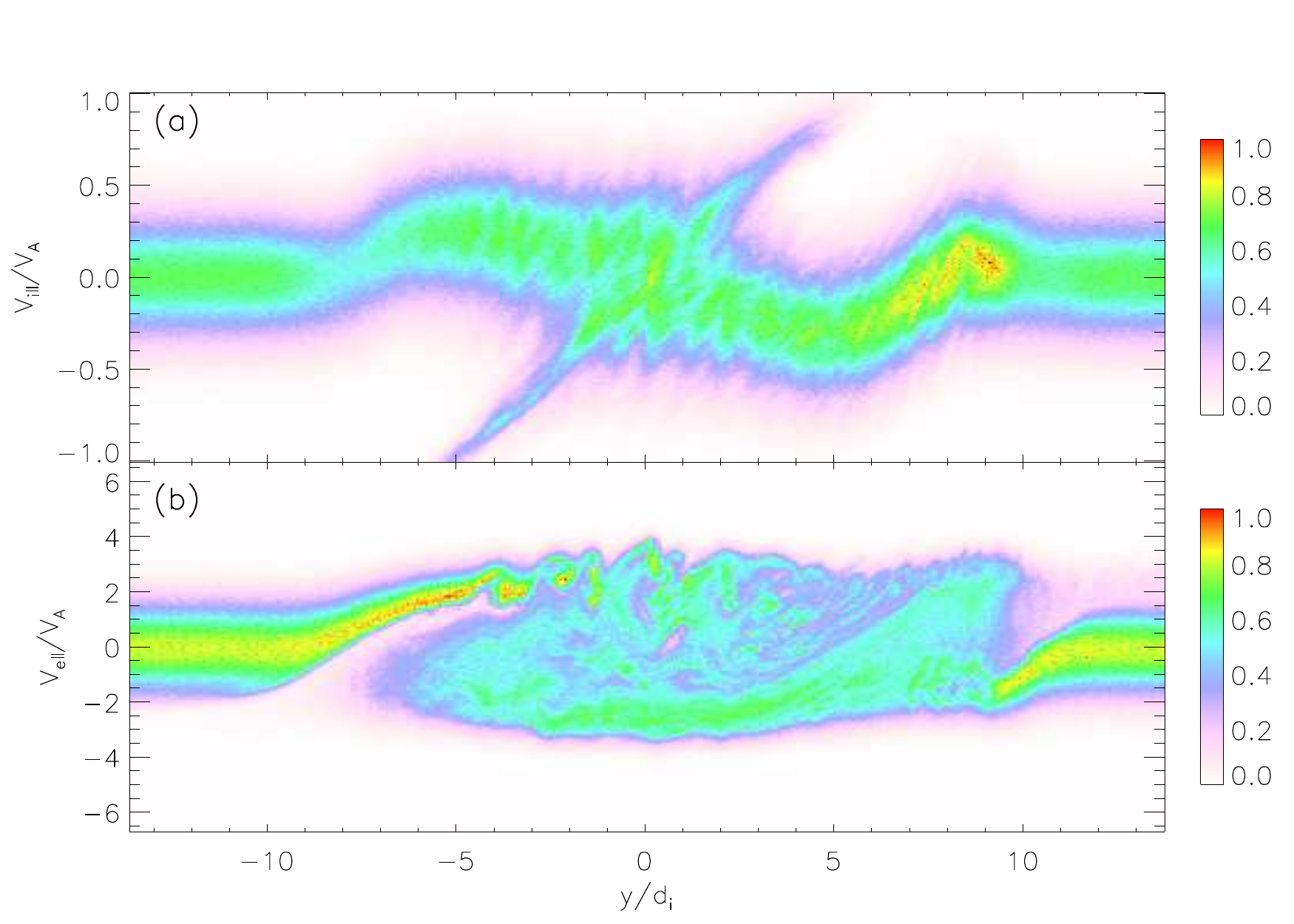}
\caption{\label{fig5} The phase space y-$V_\parallel$ for ions (a) and electrons (b) along a cut in y across the exhaust of Run 2 at x=4 and t=100. The color bar is normalized to the maximum value in each panel.}
\end{figure*}

\begin{figure*}
\includegraphics[width=\linewidth]{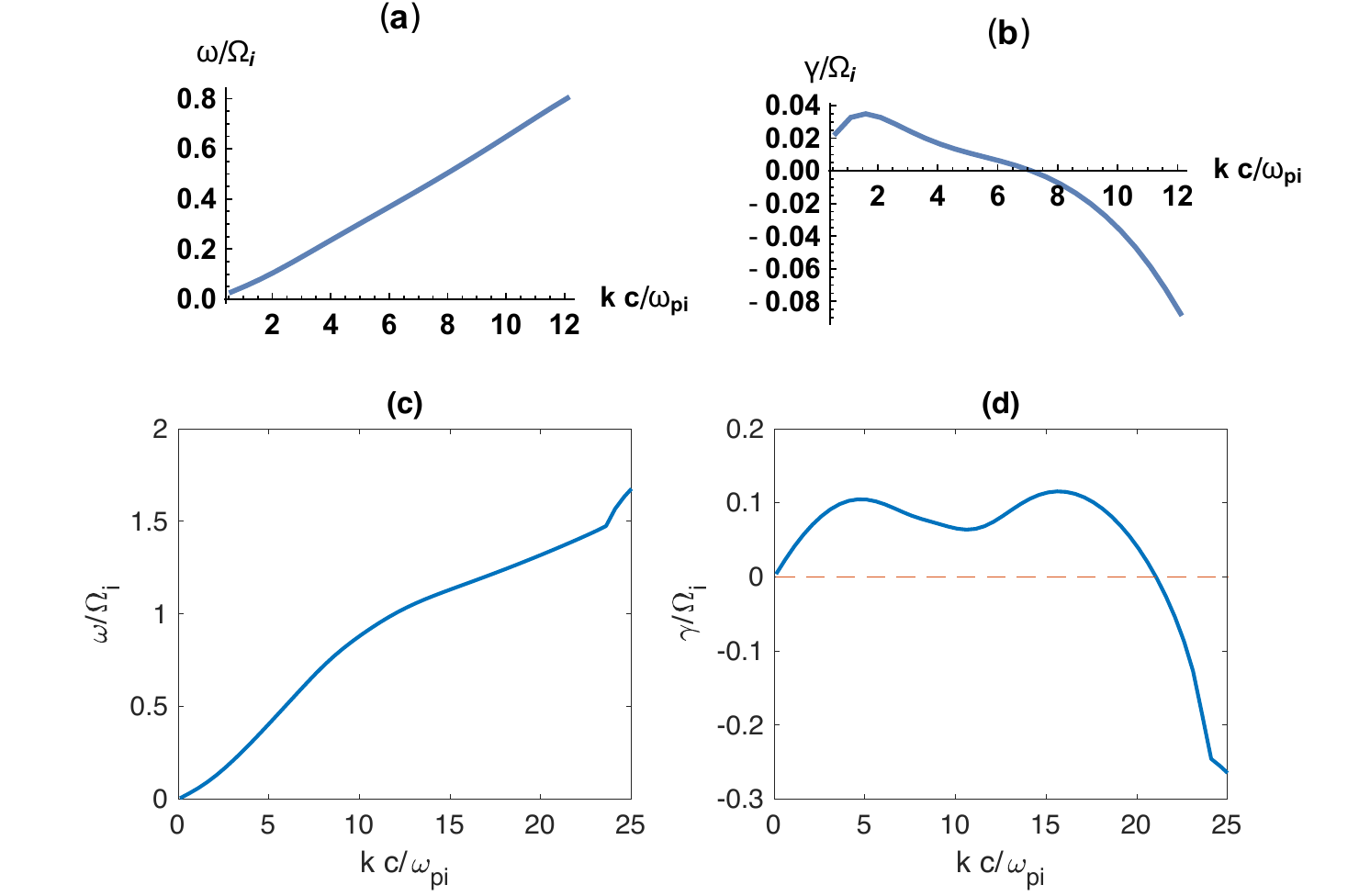}
\caption{\label{fig6} The solution $\omega$ (a) and $\gamma$ (b) versus $k$ from the dispersion relation in equation~(\ref{dispersion}). In (c) and (d), the solution of the full electromagnetic dispersion relation using the solver {\tt pdrk}\cite{Xie2016}.}
\end{figure*}

In Fig.~\ref{fig4}(e) and (f), the turbulence at the bottom RD evolves
to longer wavelength, forming well-separated, large-scale structures
that drive the electron current striations into the core of the
exhaust.  To understand why the core of the exhaust become more
turbulent further downstream and to further clarify how the striations
develop in the lower RD, we show in Fig.~\ref{fig5} the phase space
y-$V_\parallel$ for ions and electrons along a cut in y (at $x=4$ and $t=100$)
across the exhaust. We see from the electrons in panel (b) that the
instability at the cavity RD (on the left) is weak, so the electron
beam flows into the exhaust without dissipation when crossing the
RD. After being accelerated into the exhaust, the beam becomes
unstable, driving the turbulence in the exhaust. This instability has
$\omega\sim \Omega_{ci}$, $k\rho_i\sim 1$. Since the local magnetic
field is dominantly in the out-of-plane direction ($B_x\sim 0$ and
$B_z/B_y\sim 13$), this instability's in-plane wave vector is very
oblique to the magnetic field ($\cos{\theta}=0.054$). To
determine the nature of this local instability, we analyse the
electrostatic dispersion relation similar to Drummond and Rosenbluth \cite{Drummond1962} in terms of the plasma dispersion function Z as
below:
\begin{equation}
    \begin{aligned}
    \sum\limits_j\sum\limits_{n=-\infty}^\infty{\Gamma_n(k_\perp^2 \rho_j^2) \over (k \lambda_{Dj})^2}&\left\{{1\over 2} Z'\left({\omega-k_\parallel u_j-n\Omega_j \over k_\parallel v_j}\right)\right.
   \\ & \left.-{n \Omega_j \over -\omega+k_\parallel u_j+n\Omega_j}\left[1+{1 \over 2}Z'\left({\omega-k_\parallel u_j-n\Omega_j\over k_\parallel v_j}\right)\right]\right\}=1
    \end{aligned}
\end{equation}
Here $\Gamma_n(x)=e^{-x}I_n(x)$, $I_n(x)$ is the modified Bessel function of the first kind, $Z'(s)$ is the derivative of the Z function with $Z'(s)=-2(1+sZ(s))$. The subscript j stands for the different species, $v_j$ is the thermal speed, $\Omega_j$ is the cyclotron frequency, $\rho_i=v_j/\Omega_j$ is the Larmor radius, $\lambda_{Dj}=v_j/\omega_{pj}$ is the Debye length where $\omega_{pj}$ is the plasma frequency of species j, and $u_j$ is the drift speed. $k_\parallel=k \cos{\theta}$ is the parallel component of the wave vector and $k_\perp=k \sin \theta \approx k$ is the perpendicular component.
We use this dispersion relation to examine a simplified system with three populations: one electron beam (labeled as eb), one lower energy electron population (ec) and one ion population (i), which is analogous to the distributions in Fig.~\ref{fig5}(b) at around $y=-4$. For simplicity, these populations are assumed to be isotropic Maxwellian distributions and have temperatures close to their parallel temperatures in the simulation. The ion population has density one, temperature 0.05 and speed zero. The electron beam has density 0.3, temperature 0.005 and speed 2.5. The lower energy electron population has density 0.7, temperature 0.08 and speed $-2.5*0.3/0.7=-1.07$ to ensure zero current. The total magnetic field is about 1.3. We keep the $|n|\leq1$ terms in the sum over the Bessel function harmonics and use $k_\parallel v_i \ll \omega \sim\Omega_i$, $\Omega_e \gg k_\parallel v_e$ and $k_\perp^2 \rho_e^2\ll 1$ to reduce the relation to:
\begin{equation}\label{dispersion}
    \begin{aligned}
    {\Gamma_1(k_\perp^2 \rho_i^2) \over (k \lambda_{Di})^2}{\Omega_i \over \omega-\Omega_i}&-{\Gamma_1(k_\perp^2 \rho_i^2) \over (k \lambda_{Di})^2}{\Omega_i \over \omega+\Omega_i}+{1 \over 2(k \lambda_{Deb})^2}Z'\left({\omega-k_\parallel u_{eb} \over k_\parallel v_{eb}}\right)\\&+{1 \over 2(k \lambda_{Dec})^2}Z'\left({\omega-k_\parallel u_{ec} \over k_\parallel v_{ec}}\right)-{\omega_{pe}^2 \over \Omega_e^2}=1,
%    -{\rho_{eb}^2 \over  \lambda_{Deb}^2}-{ \rho_{ec}^2 \over  \lambda_{Dec}^2}
    \end{aligned}
\end{equation}
where $\omega_{pe}$ is the electron plasma frequency based on the
total electron density.  We numerically solve this dispersion relation
and plot the solution $\omega$ and $\gamma$ versus $k$ in
Fig.~\ref{fig6} (a) and (b). The growth rate reaches its maximum at
around $k=1.6$ and slowly decreases at higher $k$.  We have also used
a kinetic dispersion relation solver {\tt pdrk}\cite{Xie2016} to solve
the full electrostatic dispersion relation involving the three populations above and we get qualitatively
similar results. For the unstable modes around the maximum growth $k$,
the dominant terms in the simplified dispersion relation are the first
four terms: two ion terms, one electron beam term and one lower energy
electron term. We also tried neglecting the second ion term in the
dispersion relation (proportional to $(\omega + \Omega_i)^{-1}$) and
find that the result is qualitatively unchanged so the first ion term
(the ion cyclotron term proportional to $(\omega - \Omega_i)^{-1}$)
and the other two electron terms are dominant. These three terms
coupling together suggest that this is an oblique ion cyclotron
instability driven by the electron beam, which is similar to the instability discussed in Drummond and Rosenbluth \cite{Drummond1962}. That the instability is driven by the electron beam in these 2D simulations is also supported by the results of 3D simulations presented later in which the electron beam is suppressed and the instability is absent. In comparison to the
simulation, the $k$ value measured in the simulation is around 5,
which is several times larger than the $k$ with the maximum growth
rate from this simplified dispersion relation. This is due to the
electromagnetic effects. In fact, we have used {\tt pdrk} to solve the
full electromagnetic dispersion relation and show the results in
Fig.~\ref{fig6} (c) and (d). As seen in panel (d), there is a maximum
growth peak at around $k=5$, in good agreement with the
simulation. However, there is also another peak at around $k=15$,
which can also be found using either {\tt pdrk} electrostatic mode. However, due to their typically
higher saturation amplitudes, longer wavelength modes dominate late
time dynamics.  We conclude therefore that the striations in the
exhaust core are electromagnetic, electron-beam-driven ion cyclotron
waves.

We have used the 2D Riemann simulation as a proxy to understand the
physics of instabilities and turbulence that develop in the 2D
reconnection simulation. However, the 2D limitation forces the
instability wave vector in these simulations to be oblique to the
magnetic field and thus these simulations may not capture the true
nature of the instabilities and turbulence in a real 3D system. Yet a
3D reconnection simulation is too computationally expensive to
accommodate an exhaust extending far downstream of the x-line. We therefore 
use 3D Riemann simulations to explore the development of turbulence
in 3D reconnection exhausts.

\section{3D Riemann simulations}
\subsection{Comparison to 2D}
\begin{figure*}
\includegraphics[width=\linewidth]{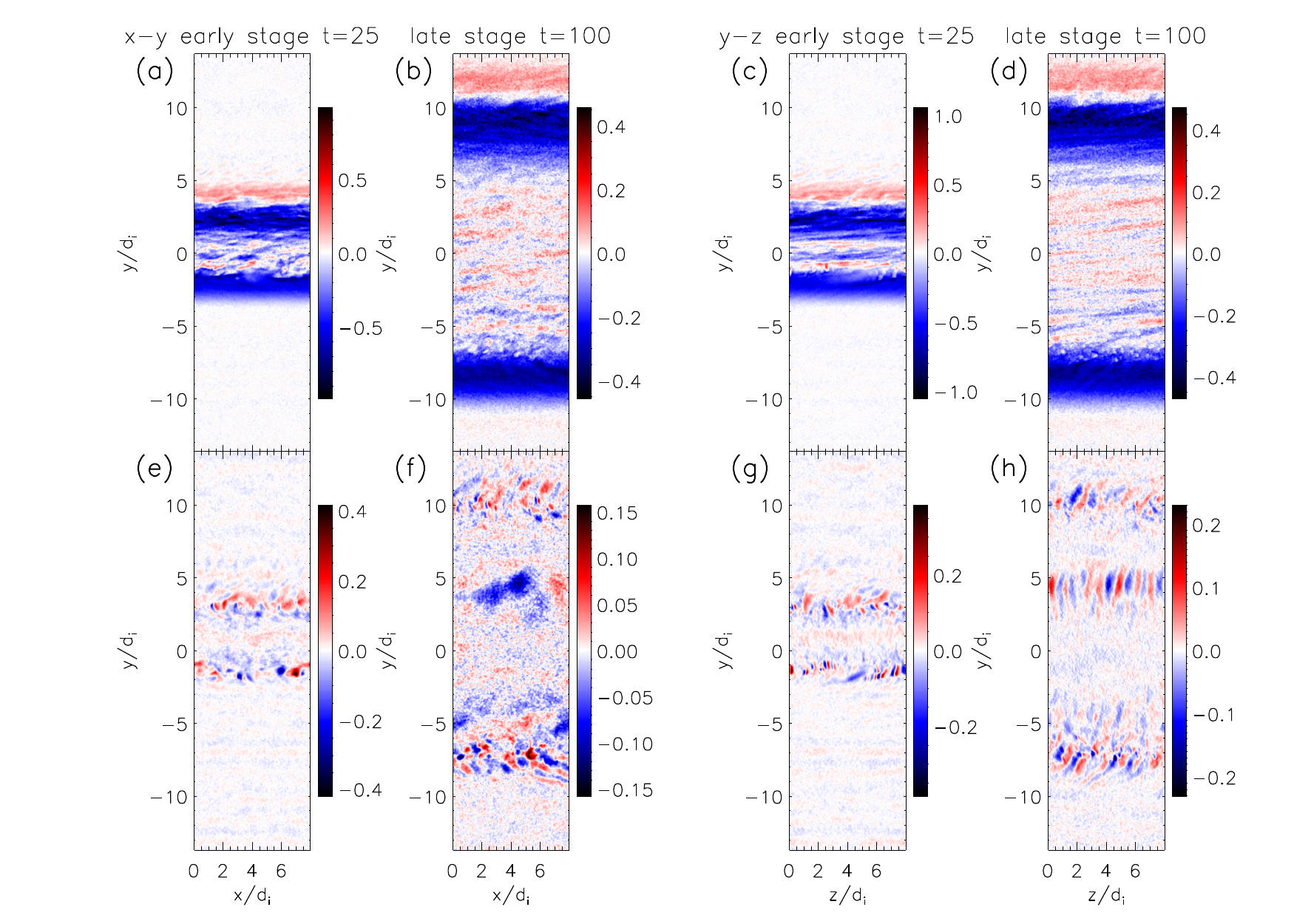}
\caption{\label{fig7} The 2D cuts (x-y and y-z) from Run 3 through the middle of the 3D domain at $z=4$ and $x=4$, respectively. In (a) and (b) x-y cuts of $J_\parallel$ and in (c) and (d) y-z cuts of $J_\parallel$. In (e) and (f) x-y cuts of $E_\parallel$ and in (g) and (h) y-z cuts of $E_\parallel$. }
\end{figure*}

In Fig.~\ref{fig7} we present the results of a 3D Riemann simulation
(Run 3) with $L_x=L_z=8, L_y=32$ and a reduced number of particles per
grid but otherwise the same physical parameters as Run 2. The data is
presented in 2D cuts (either x-y or y-z) through the middle of the 3D
domain, which is at $z=4$ and $x=4$. Due to the lower ppg and higher
noise in 3D simulations, each 2D image from 3D simulations shown in this paper is smoothed three times in succession in both dimensions with a 5-point boxcar stencil. First, the
instabilities exhibit large values of $k_z$, which could not exist in
the 2D simulation. This means that 2D simulations in the x-y plane are
not adequate to explore the dynamics of these instabilities. Within
the exhaust the magnetic field is dominantly in the z direction so the
results of Fig.~\ref{fig7} are evidence that wavevectors parallel to
the ambient magnetic field are needed to properly describe the turbulence that
develops within the exhaust. A surprise from the data in panels (a)
and (b) is that the oblique current striations within the exhaust
described previously become weak. The data in panels (g) and (h)
reveal the growth of strong instabilities with finite $k_z$ at both
RDs and within the core of the exhaust. At late time in
Fig.~\ref{fig7}(h) the instability in the bottom (low density) RD does
not weaken as it did in 2D (Fig.~\ref{fig3}(d)). The instability
within the exhaust around $y=5$ in Fig.~\ref{fig7}(h) was not present
in the 2D simulation.

\begin{figure*}
\includegraphics[width=\linewidth]{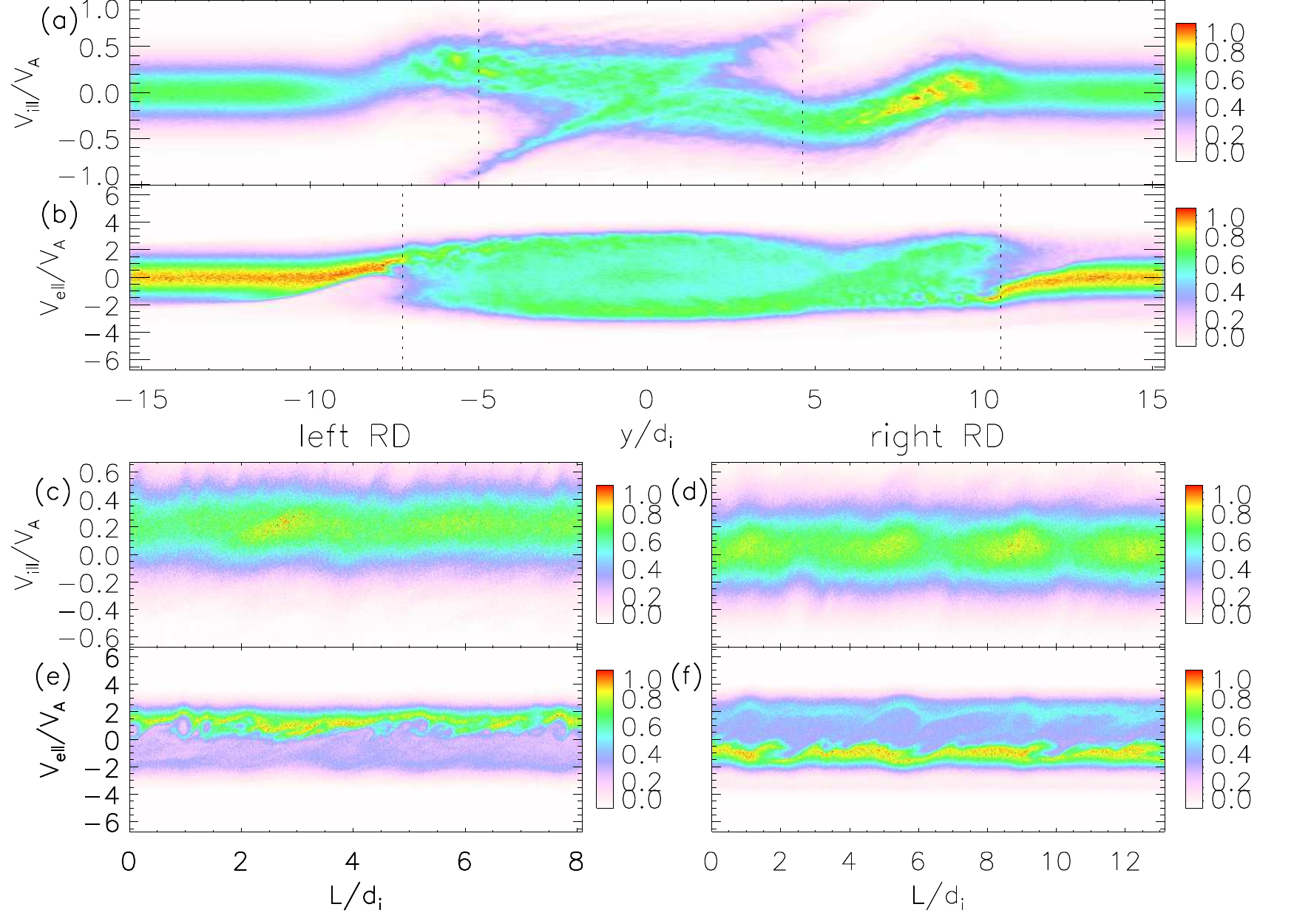}
\caption{\label{fig8} In (a) and (b) the ion and electron y-$V_\parallel$ phase space from Run 3 along a line in y across the exhaust at $x=4$, $z=4$. Dotted lines in (a) and (b) indicate the $y$ location where instabilities are peaked. The two lines in (a) are for the instability in the core of the exhaust. The other two lines in (b) are for the instabilities at both RD current sheets. The parallel phase space L-$V_\parallel$ along field lines at these two $y$ locations for ions and electrons are shown in (c)-(f), where L is the distance along the field starting from $x=0, z=0$ along the positive z direction. The color bar is normalized to the maximum value in each panel.}
\end{figure*}

The ion and electron phase spaces $y-V_\parallel$ from the Run 3
simulation at $t=100$ are shown in Fig.~\ref{fig8}. The cuts are along
a line in y across the exhaust at $x=4$, $z=4$. We plot dotted lines
in panels (a) and (b) to indicate the $y$ location where the
amplitudes of the dominant instabilities peak. The line in panel (a)
at $y=5$ marks the location of the new instability in the core of the
exhaust. It is an ion-ion streaming instability that is driven by the
counterstreaming ions that are evident in the ion phase space. The
instability is also more weakly driven around $y=-5$. This instability is strongest around the locations of the SSs. This
instability will be discussed in depth later in the context of a mass
ratio 100 simulation.  The two lines in panel (b) are for the
instabilities at the RD current sheets. We show the parallel phase
space L-$V_\parallel$ along field lines at these two $y$ locations for
ions and electrons in panels (c)-(f), where $L$ is the distance along
the magnetic field starting from $x=0, z=0$ and moving in the positive
z direction. The field line here is assumed to lie in the x-z plane, since $B_y$ is an order of magnitude smaller than the total field strength. The correlated structuring of the ion beam and two electron beams in the phase
spaces along the direction of the magnetic field show that they are a mixture of Buneman and electron-electron streaming instabilities. They result from the relative drifts between the current-supporting electron beam and the other ion or electron populations. These strong
instabilities at the low density RD dissipate the current-supporting
electron beam more efficiently than in the 2D simulation (compare
Fig.~\ref{fig5}(b) with Fig.~\ref{fig8}(b)) and prevent it from
forming the current striations in the exhaust core.

\subsection{The impact of the mass ratio on the development of turbulence}
\begin{figure*}
\includegraphics[width=\linewidth]{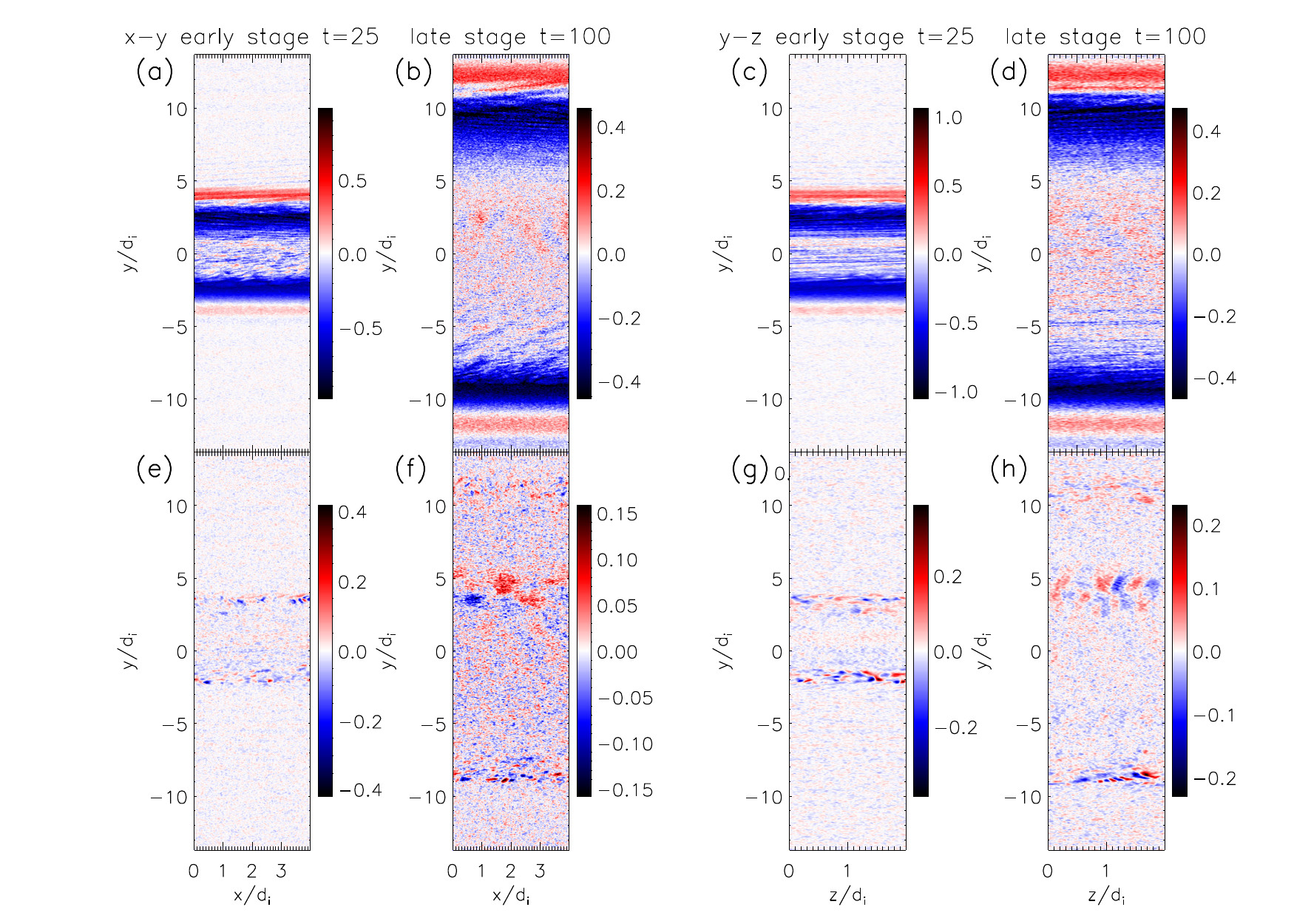}
\caption{\label{fig9} The 2D cuts (x-y and y-z) from Run 4 through the middle of the 3D domain at $z=1$ and $x=2$, respectively. In (a) and (b) x-y cuts of $J_\parallel$ and in (c) and (d) y-z cuts of $J_\parallel$. In (e) and (f) x-y cuts of $E_\parallel$ and in (g) and (h) y-z cuts of $E_\parallel$. These quantities are organized in the same way as in Fig.~\ref{fig7}, and with the same color bar for each corresponding panel. Note that we do not respect the image aspect ratio.}
\end{figure*}

The results of a 3D Riemann simulation with mass ratio 100 (Run 4)
with a domain of $L_x=4, L_z=2$ are presented in Fig.~\ref{fig9}. The
other parameters of the simulation are identical to those of Run
3. Thus, the data from the two simulations can be compared to
establish the sensitivity of turbulence drive mechanisms to the
artificial mass ratio in the simulations. The data is organized in
Fig.~\ref{fig9} in the same way as in Fig.~\ref{fig7}, and with the
same color bar for each corresponding panel. The x-y and y-z cuts are
also through the middle of the 3D domain, which is at $z=1$ and
$x=2$. Note that we do not respect the image aspect ratio here since
the the dimensions x and z would be too short to clearly display the
results. The plots displaying $E_\parallel$ reveal that the
instabilities develop at shorter wavelength in the mass-ratio 100
run. For the Buneman and electron-electron streaming instabilities at the two RDs this is consisent with
the expected scaling $k\sim\omega_{pe}/v_{de}$. The turbulence at the
RDs is also less well developed in the mass-ratio 100 run, indicating
that the turbulence is weaker. For lower electron mass the electron
thermal speed increases while the current and therefore the electron
drift speed needed to support the RD does not. Thus, the ratio of the
electron beam speed to the thermal speed is reduced in the mass-ratio
100 run. This reduces the strength of the Buneman and electron-electron streaming instabilties at the
two RDs. The electron beam speed, due to Ampere's law, is of order the Alfv\'en speed since the width of the RD is of $d_i$ scale. Therefore, in a regime where the electron thermal speed is larger than the Alfv\'en speed, or equivalently $\beta>m_e/m_i$, the instabilities at the RDs will be weak. Otherwise, when $\beta<m_e/m_i$, the instabilities at the RDs will become strong. This limit will be further studied in future papers. It is worth pointing out that if we increase the mass ratio by increasing $m_i$ instead, the Alfv\'en speed will be reduced so the ratio of the electron beam speed (scaling as the Alfv\'en speed) to the thermal speed will still be reduced. In the next section we investigate the
ion-ion instability in Run 4.
\subsection{Studies of ion-ion instabilities}
\begin{figure*}
\includegraphics[width=\linewidth]{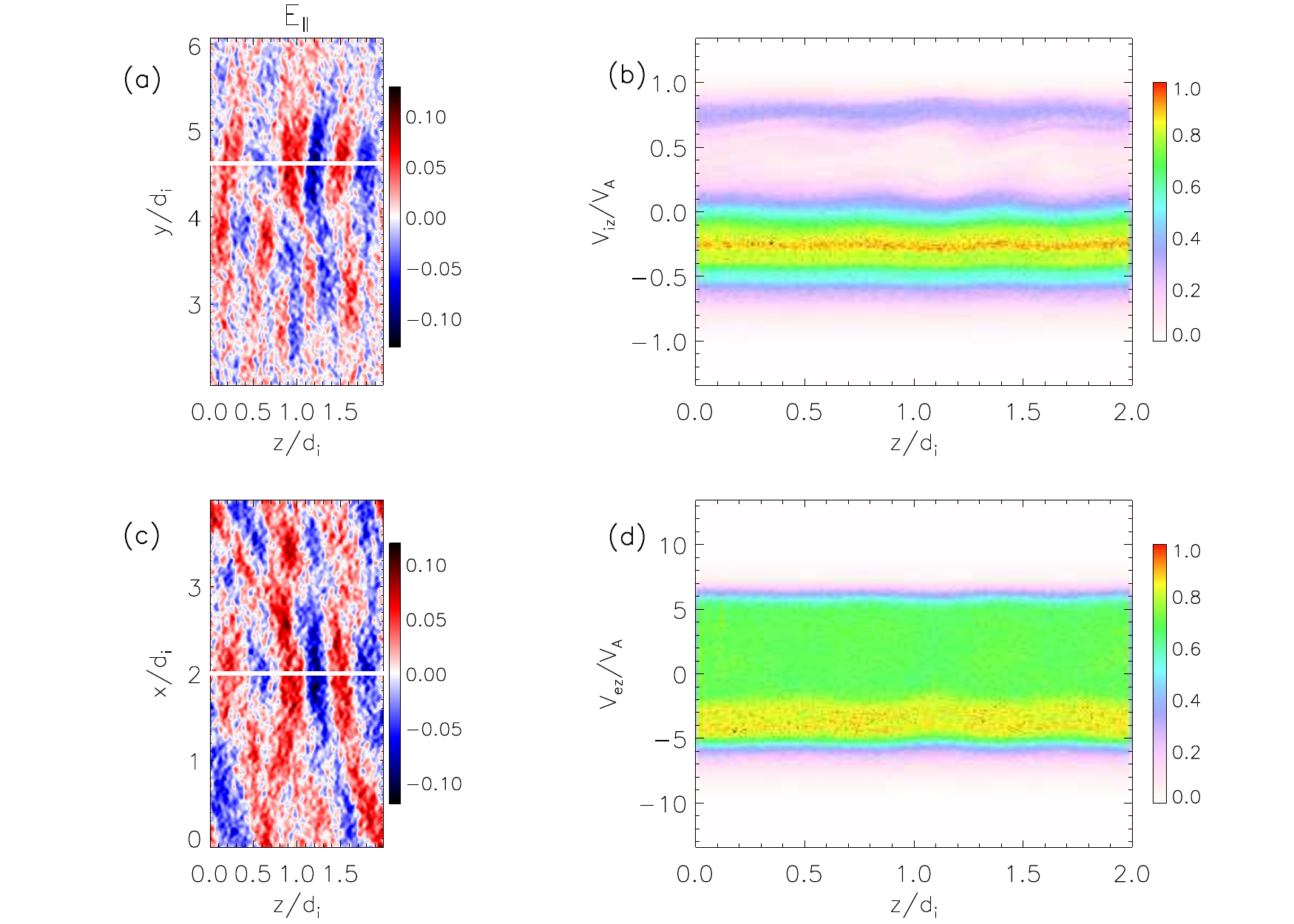}
\caption{\label{fig10} In (a) and (c) a zoom into the region around $y=5$ of Run 4.  Shown is $E_\parallel$ in y-z and x-z planes.  In (b) and (d), the ion and electron phase space z-$V_z$ along the white lines in (a) and (c). In (b) and (d), the color bar is normalized to the maximum value in each panel.}
\end{figure*}

We now explore the driver of the turbulence seen in the core of the
exhaust at the SS around $y=5$ in Figs.~\ref{fig9}(f) and (h). In
Figs.~\ref{fig10}(a) and (c) we show a blowup of $E_\parallel$ in y-z
(at $x=2$) and x-z (at $y=4.6$) planes. The white lines show the
location of the cuts. This data reveals that the dominant wavevector
is along the z direction, which is essentially the magnetic field
direction at that location. Also, given that that the magnetic field
perturbations there are weak (not shown), this is also an
electrostatic instability. In panels (b) and (d), we show the ion and
electron phase spaces z-$V_z$ along the white line in
Fig.~\ref{fig10}(c). The instability partially thermalizes the
counterstreaming ion beams and has only a small impact on electrons
so this appears to be a weak ion-ion streaming instability.

\begin{figure*}
\includegraphics[width=\linewidth]{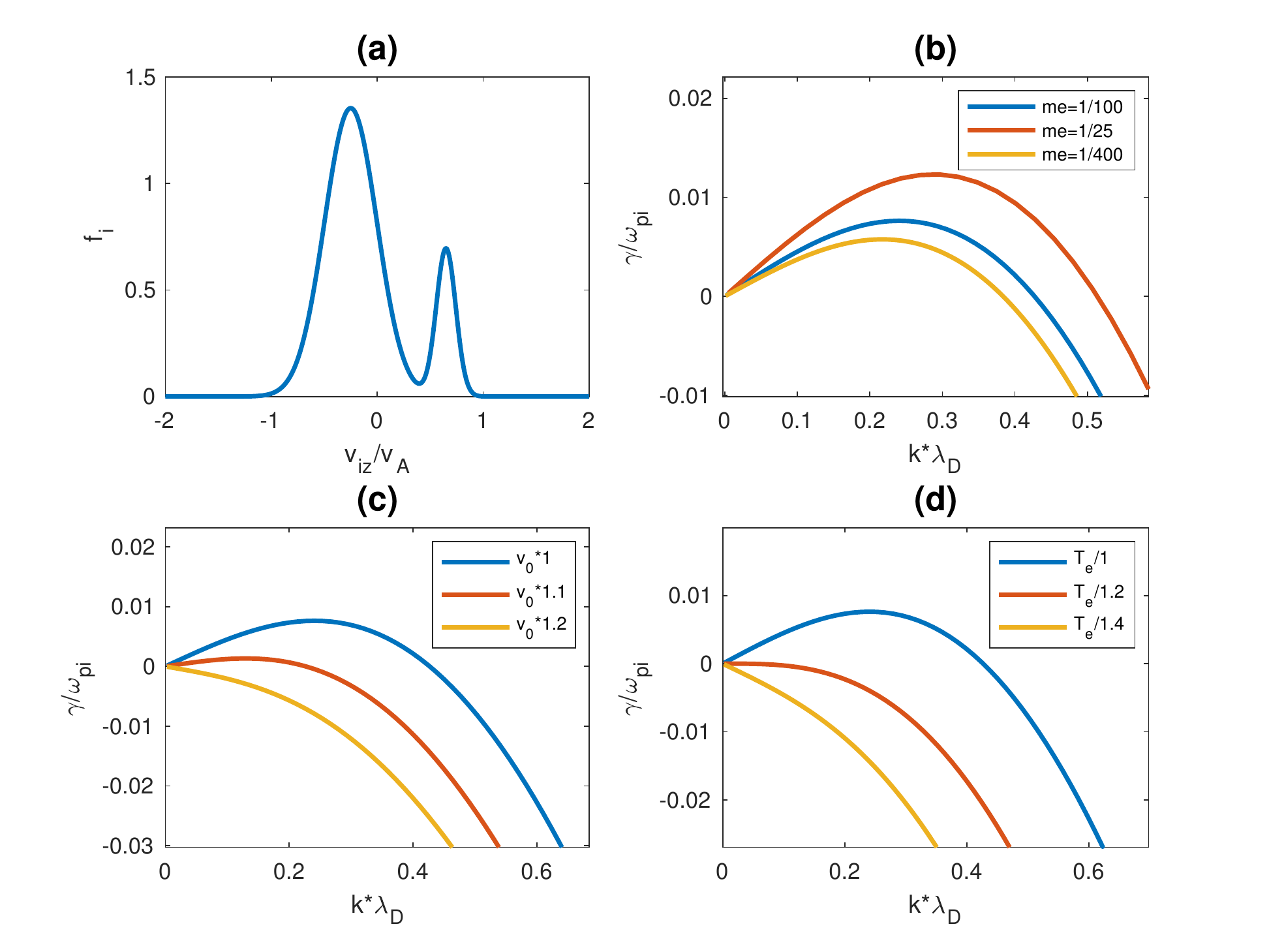}
\caption{\label{fig11} In (a) a simple model of the ion distribution prior to the ion-ion instability. The growth rates $\gamma$ are plotted versus $k$ in the blue lines in (b)-(d). The dependence of the $\gamma$  on the mass ratio, the relative drift of the two beams and the electron temperature is also shown.}
\end{figure*}

To establish this conclusion, we use {\tt pdrk}\cite{Xie2016} to show that a reasonable ion
distribution function can produce the basic characteristics of the
instability in the simulation. Specifically, we show that a reasonable
distribution function can lead to an electrostatic instability with a
maximum growth rate at the wavelength and phase speed close to that
measured in the simulation. We plot this representative distribution
function in Fig.~\ref{fig11}(a). The two beams have velocities of -0.25
and 0.65 with thermal speeds of 0.35 and 0.13. The density ratio is
84:16. By comparing to Fig.~\ref{fig10}(b), we can see that the peaks
of the beams and the thermal spread of this distribution function are
close to those in the simulation. We use a Maxwellian distribution of
electrons with an electron parallel temperature of 0.14 as measured in the
simulation. Using {\tt pdrk}, the growth
rate, $\gamma$, is plotted versus $k$ in the blue lines in panels
(b)-(d). It produces a fastest growing mode with wave length about 0.67
and phase speed about 0.41, which is comparable to that in the
simulation. Thus, the instability is driven by the counterstreaming
ions. In addition, we show the dependence of $\gamma$ on the
mass ratio, the relative drift of the two beams and the electron
temperature. As in the simulations, the speed of light is chosen to be
proportional to $1/\sqrt{m_e}$ so the electrons remain
nonrelativistic. Panel (b) reveals that the instability weakens as the mass ratio increases, which is consistent with the simulation
results. The wavelength of the fastest growing mode (normalized to the
Debye length) is not sensitive to mass ratio. If we renormalize it to
$d_i$ as in the simulations, the wave length will be roughly
proportional to $\sqrt{m_e}$, which is consistent with the
simulations. Panels (c) and (d) show that higher relative drift $V_0$ or
lower electron temperature $T_e$ have a stabilizing effect. With a small
change of these parameters, the instability becomes stable, so the
instability is close to marginal. This is consistent with the weak
disturbance of the ion phase space in Figs.~\ref{fig10}(b) and~\ref{fig10}(d). This dependence is also consistent with the theoretical results by Fujita\cite{FUJITA1977} that the instability is stabilized when the sound speed $\sqrt{T_e/m_i}$ is lower than half of the relative drift of the ion beams. In
addition, as shown in Zhang et al.\cite{Zhang2019}, the relative speed
of the counterstreaming ions in the exhaust is proportional to the
Alfv\'en speed and the electron temperature in the exhaust is of the order of the upstream temperature. Thus, lowering the upstream sound speed to Alfv\'en speed ratio (or equivalently lowering upstream $\beta$) can stabilize the instability. This is consistent with the discussion in Zhang et
al.\cite{Zhang2019} where it was concluded that the counterstreaming ion
beams would be stable in the low upstream $\beta$ limit. In consequence, the
counterstreaming ion beams that develop during low $\beta$
reconnection can propagate long distances before thermalizing and they
are therefore unable to drive the strong turbulence necessary to
produce significant dissipation at the SSs.

%fig10(c)(d) also tell us that the increasing relative beam drift and decreasing electron temperature have a strong stabilizing effect. According to reference(?), the ion beam counterstreaming speed is proportional to the magnetic field. So 

\section{Conclusion}
In this paper we have used a 2D reconnection simulation, 2D and 3D
Riemann simulations, and a kinetic dispersion relation solver to
explore the role of instabilities and turbulence in low-$\beta$,
guide-field reconnection exhausts. The overall structure of the
exhaust is controlled by a pair of RDs that bound the exhaust and
rotate the reconnecting field into the out-of-plane direction. The RDs
drive the Alfv\'enic outflow exhaust and out-of-plane flows that
propagate toward the center of the exhaust and drive a pair of slow
shocks.  The initial total plasma $\beta$ in the simulations was
chosen to be 0.1. The 2D reconnection and Riemann simulations reveal
that just downstream of the x-line the Buneman instability develops
at the low density RD, the RD in which the electron flow
supporting the magnetic rotation points towards the exhaust
center. The turbulence in this RD transitions to longer wavelength and
drives electron current striations that penetrate into the core of the
exhaust. Further downstream the entire exhaust core exhibits
large-amplitude striations that are linked to an oblique
electron-beam-driven, electromagnetic ion cyclotron instability. The
electron beam driving this instability is injected from the low
density RD. However, in 3D Riemann simulations, the
additional dimension enables the growth of strong Buneman and
electron-electron streaming instabilities at both RDs by allowing
a non-zero $k_z$, which is along the strong guide field $B_z$. These strong
instabilities suppress the generation of the electron beam at the low
density RD with the consequence that the current striations in the
core of the exhaust are largely suppressed.  However, the instabilities at the RDs
become weaker with higher ion-to-electron mass ratio due to the higher
electron thermal speed compared with the electron beam speed (of the order
of the Alfv\'en speed). In the regime of $\beta > m_e/m_i$, the electron thermal
speed is larger than the Alfv\'en speed and these instabilities will
be weak. Otherwise they will be strong. The strong turbulence regime
of $\beta < m_e/m_i$ will be further explored in the future.  In
addition to the Buneman and electron-electron streaming instabilities
at the two RD current sheets, the 3D simulation also reveals an ion-ion streaming instability in the core of the exhaust at the SSs which acts to partially thermalize the counterstreaming ion beams that generate the SSs.
%Thus, in a system with real mass ratio, it is unclear what impact the the Buneman instability at the RDs will be. Note that we are in the regime of $\beta > m_e/m_i$ where $V_eth>V_A$. 
We use a kinetic dispersion relation solver to show the ion streaming
instability is stable when the Alfv\'en speed greatly exceeds the ion
sound speed at low upstream $\beta$. The consequence is the
counterstreaming ion beams that develop during low $\beta$
reconnection can propagate long distances before thermalizing and they
are therefore incapable of driving the strong turbulence necessary to
produce significant dissipation at the SSs. The direct exploration of
the very low $\beta$ regime to further establish the stability of
the counterstreaming ion beams is a topic for future work.

The results suggest that in realistic low-$\beta$ guide field
reconnection exhausts with upstream $\beta>m_e/m_i$, the slow shocks
inside the exhaust are largely laminar and the instabilities and
turbulence that develop will likely be too weak to play a significant
role in energy conversion. Without strong turbulence at the slow
shocks, neither species can undergo the canonical diffusive shock
acceleration. Energy conversion in the exhaust will be dominantly
controlled by laminar physics with little dissipation or scattering
from turbulence. The results are therefore consistent with the
conclusions by Zhang et al.\cite{Zhang2019}. This conclusion has broad
implications for understanding reconnection in the solar corona and
the inner heliosphere. The types and role of the instabilities and
turbulence can be tested from the data expected from the Parker Solar Probe
\cite{Bale2016,Kasper2016} as it approaches the low-$\beta$
environment in the outer reaches of the solar corona.

\begin{acknowledgments}
This work was supported by NSF Grant Nos. PHY1805829 and PHY1500460,
NASA Grant Nos. NNX14AC78G and NNX17AG27G, and the FIELDS team of the
Parker Solar Probe (NASA Contract No. NNN06AA01C).  Simulations were
carried out at the National Energy Research Scientific Computing
Center. Simulation data are available on request.
\end{acknowledgments}

\bibliography{reference}

\newpage

%\onecolumngrid
\begin{widetext}
\begin{table}[ht]
\caption{Simulation parameters} % title of Table
\centering % used for centering table
\begin{tabular}{c c c c c c c c c c c} % centered columns  (4 columns)
\hline\hline %inserts double horizontal lines
Run  & $m_i/m_e$ & $B_{x,a}$ & $B_{z,a}$ & $T_i=T_e$ &  dims & $L_y \times L_x \times L_z$ &$c^2$&dx&dt&ppg\\ [0.5ex] % inserts table
%heading
\hline % inserts single horizontal line
1 & 25 & 1 & 1 & 0.05&2&102.4$\times$409.6 $\times$ 0&$45$&0.0125&5.9e-3&100 \\ % inserting body of the table
2 & 25 & 1 & 1 & 0.05&2&102.4$\times$16$\times$0&$45$&0.0125&5.9e-3&100 \\
3 & 25 & 1 & 1 & 0.05&3&32$\times$8$\times$8 &$45$&0.0125&5.9e-3&25\\
4 & 100 & 1 & 1 & 0.05&3&32$\times$4$\times$2 &$180$&0.00625&2.95e-3&25\\[1ex] % [1ex] adds vertical space
\hline %inserts single line
\end{tabular}
\label{table1} % is used to refer this table in the text
\end{table}
\end{widetext}



\end{document}